\documentstyle[aps]{revtex}
\begin{document}
\draft
\tighten
\title{BREMSSTRAHLUNG PAIR PRODUCTION IN RELATIVISTIC HEAVY ION COLLISION}
\author{Helmar Meier$^1$, Kai Hencken$^{1,2}$, Dirk Trautmann$^1$,
         and Gerhard Baur$^3$}
\address{$^1$Institut f\"ur Physik der Universit\"at Basel,
          4056 Basel, Switzerland \\
          $^2$National Institute of Nuclear Theory, University of Washington, 
          Seattle, WA 98195, USA \\
          $^3$Institut f\"ur Kernphysik (Theorie), Forschungszentrum J\"ulich, 
          52425 J\"ulich, Germany}
\date{\today}
\maketitle
\begin{abstract}
We calculate production of electron- and muon-pairs by the
bremsstrahlung process in hadron collisions
and compare it with the dominant two-photon
process. Results for the total cross section are given for
proton-proton and heavy-ion collisions at energies of the Relativistic
Heavy Ion Collider (RHIC) and the Large Hadron Collider (LHC).
\end{abstract}
\section{Introduction}
Dileptons produced in central heavy ion collisions are investigated as
one of the possible signal for the formation of the quark gluon
plasma. The electromagnetic production of these leptons mainly in
peripheral collisions is the dominant background process and therefore
has to be studied in detail. Furthermore the production of muon pairs
has been proposed as a possible way to measure the beam luminosity in
the storage ring at the Large Hadron Collider (LHC) \cite{eggert}.
 In order to be comparable
with other methods the rate of muon pair production has to be known to
better than one percent. A recent study shows \cite{conradt}, that
inelastic photon emission processes, where the proton is excited to,
e.g., a $\Delta$, already contribute at this level. In this article we
want to study the importance of the bremsstrahlung process for the
total rate.

Counting powers of $Z$ for different processes, one finds
that the bremsstrahlung process is proportional to $Z^6$, that is, two
powers of $Z$ enhanced compared to the dominant two-photon process
($\sim Z^4$). Therefore one could expect this process to be at least
of the same order of magnitude as the two-photon process for very
heavy ions. But for high energies it was found to be much smaller. An
estimate of the cross section was done in \cite{ginzburg} and a more
detailed calculation for $e^-e^+$ collisions at LEPII energies can be
found in \cite{panella}. The two processes lead to lepton pairs with 
different $C$-parity. Therefore the interference between the two 
mechanism vanishes for the total cross section.
It was found that only for large scattering angles of the projectile
the bremsstrahlung production is dominant (see \cite{brodsky} for a
calculation of pair production from muon-nucleus collisions). Large
scattering angles do not contribute much to the total cross sections
and for nuclei they are further reduced due to the finite
size. Collisions with large momentum transfer are very likely to breakup 
the nucleus; the elastic form factor falls off quickly for large $q^2$.

\section{lepton pair production}
The cross section for the two-photon pair production has been
calculated in the semi-classical approximation in \cite{alscher},
where the fields of the colliding ions have been treated as external
fields using also the straight line approximation. Small impact
parameters, smaller than twice the nuclear radius, where the ions
interact strongly, have been cut off. Whereas the analysis of
\cite{alscher} was mainly on studying the production of multiple
(electron-positron) pairs, here we are considering only single-pair creation.

The cross section for the bremsstrahlung pair production has been calculated
in lowest order perturbation theory \cite{bjorken,landau}.
One of the corresponding Feynman diagrams is shown in Fig.\ \ref{brems-graph}.

\begin{figure}
\caption{A Feynman diagram of bremsstrahlung pair production.}
\label{brems-graph}
\end{figure}

Using 'Heaviside-convention', that is,
\begin{equation}
\alpha=\frac{e^2}{4\pi\hbar c}
\end{equation}
the S-matrix element is given by
\begin{eqnarray}
S_{fi} & = & i\frac{Z_A^2Z_Be^4}{V^3}\sqrt{\frac{M_A^2M_B^2m^2}{E_f^AE_i^AE_f^B
E_i^BE^{e^-}E^{e^+}}}\nonumber \\
& & \times(2\pi)^4\delta^4(p_f^A+p_f^B+p^{e^-}+p^{e^+}-p_i^A-p_i^B)\nonumber \\
& & \times\{\frac{1}{(p^{e^-}+p^{e^+})^2+i\varepsilon}\quad\frac{1}
{(p_i^B-p_f^B)^2+i\varepsilon}\nonumber \\
& & \times\ [\overline u (p_f^A,s_f^A)\gamma_{\mu}\frac{1}{p\hspace{-.4em}/_f^A+
p\hspace{-.4em}/^{e^-}+p\hspace{-.4em}/^{e^+}
-M_A+i\varepsilon}\gamma_{\nu}u(p_i^A,s_i^A)\nonumber \\
& & \times\ \,\,\overline u(p^{e^-},s^{e^-})\gamma^{\mu}v(p^{e^+},s^{e^+})
\ \times\ \overline u(p_f^B,s_f^B)\gamma^{\nu}u(p_i^B,s_i^B)\nonumber \\
& & +\ \,\,\overline u (p_f^A,s_f^A)\gamma_{\mu}
\frac{1}{p\hspace{-.4em}/_f^A+p\hspace{-.4em}/_f^B-
p\hspace{-.4em}/_i^B-M_A+i\varepsilon}\gamma_{\nu}u(p_i^A,s_i^A)\nonumber \\
& & \times\ \,\,\overline u(p_f^B,s_f^B)\gamma^{\mu}u(p_i^B,s_i^B)\ \times
\ \overline u(p^{e^-},s^{e^-})\gamma^{\nu}v(p^{e^+},s^{e^+})]\},
\label{sfi}
\end{eqnarray}
where 'A' and 'B' denotes one of the colliding nuclei.
'$e^+$' and '$e^-$' denote the produced leptons, 
'$u(p,s)$' and '$v(p,s)$' are
the usual Dirac spinors for the plane wave solution.
Describing the heavy ions as spin-$1/2$ particles corresponds
to giving them a magnetic moment corresponding to $g=2$, much larger
than their experimental value. But due to their large mass, spin
dependent terms, which are of order $1/M$, are effectively
suppressed. Therefore our calculation is really independent of the
real spin of the nuclei and is valid also for spin-0 nuclei. We are
taking the finite size of the two nuclei into account by using
elastic form factors (eqs.~(\ref{ffp} and (\ref{ffhi})). 
The cross section is given by
\begin{eqnarray}
\sigma & = & \int\frac{M_AM_B}{\sqrt{(p_i^Ap_i^B)^2-M_A^2M_B^2}}
\mid \overline M_{fi}\mid^2\nonumber \\
& & \times(2\pi)^4\delta^4(p_f^A+p_f^B+p^{e^-}+p^{e^+}-p_i^A-p_i^B)\nonumber \\
& & \times\frac{M_Ad^3p_f^A}{(2\pi)^3E_f^A}
\frac{M_Bd^3p_f^B}{(2\pi)^3E_f^B}\frac
{md^3p^{e^-}}{(2\pi)^3E^{e^-}}\frac{md^3p^{e^+}}{(2\pi)^3E^{e^+}}
\label{cs}
\end{eqnarray}
with the matrix element squared
\widetext
\begin{eqnarray}
&&\hspace{-2.5em}\mid \overline M_{fi}
\mid^2\ =\ \frac{1}{4}Z_A^4Z_B^2e^8\frac{1}
{(p^{e^-}+p^{e^+})^4}\:\frac{1}{(p_i^B-p_f^B)^4}\nonumber \\
&&\hspace{-2.5em}\times\:\mbox{Tr} \frac 
{p\hspace{-.4em}/ ^{e^-}+m}{2m}\gamma_{\mu}\frac
{p\hspace{-.4em}/^{e^+}+m}{2m}\gamma_{\sigma}\nonumber \\
&&\hspace{-2.5em}\times\:\mbox{Tr}\frac{p\hspace
{-.4em}/_f^B+M_B}{2M_f^B}\gamma_{\nu}\frac
{p\hspace{-.4em}/_i^B+M_B}{2M_B}\gamma_{\lambda}\nonumber \\
&&\hspace{-2.5em}\times(\mbox{Tr}\frac{p\hspace
{-.4em}/_f^A+M_A}{2M_A}\gamma^{\mu}\frac
{p\hspace{-.4em}/_f^A+p\hspace{-.4em}/^{e^-}
+p\hspace{-.4em}/^{e^+}+M_A}{(p_f^A+p^{e^-}+p^{e^+})^2-M_A^2}
\gamma^{\nu}\frac{p\hspace{-.4em}/_i^A+M_A}{2M_A}
\gamma^{\lambda}\frac{p\hspace{-.4em}/_f^A+p\hspace{-.4em}/
^{e^-}+p\hspace{-.4em}/^{e^+}+M_A}
{(p_f^A+p^{e^-}+p^{e^+})^2-M_A^2}\gamma^{\sigma}\nonumber \\
&&\hspace{-2.5em}+\:\mbox{Tr}\frac{p\hspace{-.4em}/_f^A+M_A}
{2M_A}\gamma^{\mu}\frac
{p\hspace{-.4em}/_f^A+p\hspace{-.4em}/^{e^-}+p
\hspace{-.4em}/^{e^+}+M_A}{(p_f^A+p^{e^-}+p^{e^+})^2-M_A^2}
\gamma^{\nu}
\frac{p\hspace{-.4em}/_i^A-M_A^2}{2M_A}\gamma^{\sigma}
\frac{p\hspace{-.4em}/_f^A+p\hspace{-.4em}/_f^B-
p\hspace{-.4em}/_i^B+M_A}{(p_f^A+p_f^B-p_i^B)^2-M_A^2}
\gamma^{\lambda}\nonumber \\
&&\hspace{-2.5em}+\:\mbox{Tr}\frac{p\hspace{-.4em}/_f^A+M_A}
{2M_A}\gamma^{\nu}\frac
{p\hspace{-.4em}/_f^A+p\hspace{-.4em}/_f^B-p\hspace{-.4em}/_i^B
+M_A}{(p_f^A+p_f^B-p_i^B)^2-M_A^2}\gamma^{\mu}
\frac{p\hspace{-.4em}/_i^A+M_A}{2M_A}\gamma^{\lambda}
\frac{p\hspace{-.4em}/_f^A+p\hspace{-.4em}/^{e^-}+p\hspace{-.4em}/
^{e^+}+M_A}
{(p_f^A+p^{e^-}+p^{e^+})^2-M_A^2}\gamma^{\sigma}\nonumber \\
&&\hspace{-2.5em}+\:\mbox{Tr}\frac{p\hspace{-.4em}/_f^A+M_A}
{2M_A}\gamma^{\nu}\frac
{p\hspace{-.4em}/_f^A+p\hspace{-.4em}/_f^B-p
\hspace{-.4em}/_i^B+M_A}{(p_f^A+p_f^B-p_i^B)^2-M_A^2}\gamma^{\mu}
\frac{p\hspace{-.4em}/_i^A+M_A}{2M_A}\gamma^{\sigma}
\frac{p\hspace{-.4em}/_f^A+p\hspace{-.4em}/_f^B-p\hspace{-.4em}/_i^B+
M_A}{(p_f^A+p_f^B-p_i^B)^2-M_A^2}\gamma^{\lambda})\ .
\label{mfi}
\end{eqnarray} 
In eq.~(\ref{sfi}) and~(\ref{mfi}) we only consider the case, where 
ion 'A' emits the virtual (space-like) photon, that creates the pair.
To this we have to add the diagrams, where the virtual photon is
emitted by ion 'B'. Furthermore as the ions are identical, we have to
add the contribution from the exchange diagrams. At high energies the
interference terms between these different processes are completely
negligible. In order to be significant they must have a sufficient 
overlap in the final phase space. We can estimate their importance by
comparing the product of the absolute value of the
matrix elements with the cross section of each of them. 
This estimation shows that we can neglect all interference 
effects. We get the total cross section by multiplying eq.~(\ref{cs}) by a
factor of two, which accounts for the photon emission from ion 'B'.
We calculate the trace in eq.~(\ref{mfi}) with the help of the
algebraical calculation program {\sc form} \cite{vermaseren}.

\section{The form factors}
As mentioned in the previous section, we have to take into account
the finite size of the nuclei. We introduce a form factor at each
vertex in the Feynman diagram. This changes the equation of the
absolute value squared of the matrix element in in eq.~(\ref{mfi}) to
\begin{equation}
\mid \overline{M}_{fi}\mid^2\rightarrow\mid \overline{M}_{fi}\mid^2\mid 
F_A(q_1^2)\mid^2 \mid F_B(q_1^2)\mid^2 \mid F_T(q_2^2)\mid^2\:.
\label{nmfi}
\end{equation}
We need both space-like and time-like form factors for the nuclei.
'$q_1$' is the momentum transferred between nuclei 'A'
and 'B' and is therefor space-like ($q_1^2<0$). '$q_2$' is the
momentum of the virtual photon creating the pair; as $q_2^2$ is
identical to the invariant mass of the lepton pair, it has to be at
least $4\ m_{lepton}^2$. Therefore it is time-like ($q_2^2>0$).

In the space-like region we use the usual dipole form factor
for proton-proton collisions 
\begin{equation}
F_p(q_1^2)=\left(\frac1{1+\frac{\mid q_1^2\mid}{\Lambda^2}}\right)^2\ \ \ \ \ ;
\Lambda^2=0.71\ \mbox{GeV}^2
\label{ffp}
\end{equation}
and for heavy ion collisions a Gaussian form factor
\begin{equation}
F_{h.i.}(q_1^2)\ =\ \exp\left(-\frac{\mid q_1^2\mid}{\lambda^2}\right)
\label{ffhi}
\end{equation}
with $\lambda^2=\frac{6}{\langle r^2\rangle}$ and $\sqrt{\langle r^2\rangle}
=1.0\:A^{\frac{1}{3}}\:fm $, where $A$ is the mass number of the nucleus. 
The additional effect of the elastic form
factor is that they suppress collisions with a large $q$-value,
corresponding to small impact parameter $b$. Therefore we do not
correct for impact parameter smaller than $2R$, where both nuclei
interact strongly. This effect could be included in principle within a
Glauber model approach but is technically rather involved. In addition
our calculation (neglecting this effect), gives an upper limit of the
true cross section.

In the time-like region we make use of vector meson dominance
(VMD), that is, we assume that the photon couple to the hadrons
through vector mesons only. 
We use a form factor $F_{VMD}$ as discussed in
\cite{mosel}. As done there, we neglect the dependence 
due to the fact that the intermediate nucleus is off the mass-shell.
We do not sum over all intermediate states, taking only the
ground state as intermediate state into account. In addition we don't
take the absorption of the meson inside the nucleus into account. Both
effects reduce the total cross section \cite{titov}, therefore our calculation is
again an upper limit.
\begin{equation}
F_T(q_2^2)\ \approx\ F_{VMD}(q_2^2)\ =\ \frac{m_{V}^2}{m_{V}^2-q_2^2}
\label{fvmd}
\end{equation}
Here $V$ is one the different light vector mesons ($\rho$-, $\omega$- or 
$\phi$-meson) through which the photon couples to the nucleus.
In our calculations we only include the VMD form factor for the $\omega$-meson.
The complex mass $m_{V}$ is then given by
\begin{equation}
m_{V}\ =\ m_{\omega}\ -\ \frac{i\Gamma_{\omega}}2
\end{equation}
with $m_{\omega}\ =\ 782MeV$ is the $\omega$-meson and 
$\Gamma_{\omega}\ =\ 8.4MeV$ is
its full decay width \cite{booklet}.

Due to this form of the form factor the differential cross section of
the bremsstrahlung process has a very narrow resonance-like peak for
invariant masses near the $\omega$ mass. As the form factor 
assume that the vector mesons couples coherently with the whole
nucleus, this form factor is an upper limit for the VMD contribution.
For heavy ions it is very unlikely that the vector meson couples
coherently to all the nucleons due to their large size. Therefore we
make use of the VMD only for proton-proton collisions, neglecting it
completely for the heavy ion case. Using on the other hand the cross
section as a function of the invariant mass (see Figs. \ref{gpem_2}
and~\ref{gpmm_2}) it is easy to calculate the total cross section for
any realistic form factor.

We treat the two cases of electron-positron and muon-antimuon pair
production separately. For electron-positron pair production the differential
cross section $\left(=\frac{d\sigma}{dM_{\gamma}}\right)$ for invariant
masses equal to vector meson masses is too small to contribute
significantly to the total cross section (see Fig.\ \ref{bkpem_2} and
Fig.\ \ref{gpem_2}). In contrast to this they are important for the
muon-antimuon pair production as their mass is comparable to them.

\section{Results and conclusions}
The total cross sections are obtained from eq.~(\ref{cs}) by a Monte Carlo
integration over all possible final states ({\sc vegas} \cite{lepage}).
As the phase-space for this four particle final state is rather restricted, 
we have chosen integration variable so that we integrate only over the 
allowed (physical) phase-space using the variables given in \cite{byckling}.
For this we split the four-particle phase-space into a chain of
successive two particle decays integrating over the invariant mass of
the intermediate states and the angles in the center of mass of the
decaying particle. Momenta therefore have to be boosted from one
center of mass system to the next one. Using these adequate phase space
variables, the Monte Carlo integration can be done effectively.

\subsection{Total cross sections}
We give the total cross sections for the bremsstrahlung
production and the corresponding two-photon process for
electron-positron $(\sigma_{e^-e^+})$ and muon-pairs
($\sigma_{\mu^-\mu^+}$)
for RHIC and LHC energies in tables~\ref{brems-tpp_e}
and~\ref{brems-tpp_m}. The beam energy is giving in terms of the
Lorentz gamma factor $E=\gamma\ M$ in the collider reference frame
(center of mass of the two ions). We have calculated results for
different ion species: $^{40}_{20}$Ca, $^{208}_{82}$Pb and $^{197}_{79}$Au. 
As described above the form factor ($F_{VMD}$) is
used in the calculation for the results marked as 'Proton (VMD)'.

The total cross sections of bremsstrahlung production are compared with the
results for two-photon production, also shown
in tables~\ref{brems-tpp_e} and~\ref{brems-tpp_m}.

\subsection{Differential cross sections of bremsstrahlung pair production}

Figures~\ref{bkpem_2} to~\ref{gpmt} show the results for the
differential cross sections as a function of the invariant mass of the
produced pair ($\frac{d\sigma}{dM_{\gamma}}$) and also as a function of
the angle with respect to the beam ($\frac{d\sigma}{d\theta}$). We do
not show results using the VMD form factor. The effect of this form
factor can be easily found by multiplying the cross section with the 
absolute value square of it (see Eqs.~(\ref{nmfi}) and~(\ref{fvmd})). 
It results in a peaks around $q_2^2\approx m_\omega^2$ in the
differential cross section $\frac{d\sigma}{dM_{\gamma}}$.
\begin{figure}
\caption{The cross section $\sigma$ as a function of the  
mass of the virtual photon $M_{\gamma}$ $\left(=\frac{d\sigma}
{dM_{\gamma}}\right)$ which creates the electron-positron pair. 
The solid line are the results for a lead beam, the dot-dashed 
line for a calcium beam and the dashed line for a proton beam 
at LHC energies.}
\label{bkpem_2}
\end{figure}
\begin{figure}
\caption{The cross section $\sigma$ as a function of the 
mass of the virtual photon $M_{\gamma}$ which creates
the electron-positron pair $\left(=\frac{d\sigma}{dM_{\gamma}}\right)$. 
The solid line are the results for a gold beam 
and the dashed line for a proton beam at RHIC energies.}
\label{gpem_2}
\end{figure}
\begin{figure}
\caption{The cross section $\sigma$ as a function of the  
scattering angle $\theta$ of the created electron (or positron) 
with the beam axis $\left(=\frac{d\sigma}{d\theta}\right)$.
The solid line are the results for a lead beam, the dot-dashed line 
for a calcium beam and the dashed line for a proton beam at LHC energies.}
\label{bkpet}
\end{figure}
\begin{figure}
\caption{The cross section $\sigma$ as a function of the  
scattering angle $\theta$ of the created 
electron (or positron) with the beam axis $\left(=\frac{d\sigma}{d\theta}
\right)$. The solid line are the results for a gold beam
and the dashed line for a proton beam at RHIC energies.}
\label{gpet}
\end{figure}
\begin{figure}
\caption{same as Fig. 2 but for muon-antimuon pair production}
\label{bkpmm_2}
\end{figure}
\begin{figure}
\caption{same as Fig. 3 but for muon-antimuon pair production}
\label{gpmm_2}
\end{figure}
\begin{figure}
\caption{same as Fig. 4 but for muon-antimuon pair production}
\label{bkpmt}
\end{figure}
\begin{figure}
\caption{same as Fig. 5 but for muon-antimuon pair production}
\label{gpmt}
\end{figure}

\subsection{Conclusions}
Comparing the results for bremsstrahlung and two-photon production in 
tables~\ref{brems-tpp_e} and~\ref{brems-tpp_m}, we see that the
bremsstrahlung production contributes only by $< 3\times 10^{-3}$
to the rate of the created pairs by electromagnetic processes in heavy ion 
collision. Please have also in mind that our calculation is an upper 
limit of this cross section, therefore the real cross section is expected 
to be even smaller.
So the additional factor $Z^2$ in the cross section cannot
make up for the almost divergent photon propagators in two-photon case.
The matrix elements of the two-photon process 
contain denominators which are very
close to zero. This corresponds to the 
exchange of quasi real photons and does not 
occur for the bremsstrahlung case.
The differential cross section for bremsstrahlung production without
using the VMD form factor is also small compared to the dominant
two-photon case. If one includes the VMD form factor, one get a 
peak in $\frac{d\sigma}{dM_{\gamma}}$ around the $\omega$-meson mass. 
The peaks are absent in the two-photon case. This could be
used as a way to get more insight into the VMD contribution to this
process and especially the time-like form factor of the proton and of nuclei.
With respect to the use of muon pairs as a luminosity monitor
for the LHC, we see that the bremsstrahlung production seems to be too
small to be of importance at the 1\% level. The largest uncertainty in
our calculations is the almost unknown time-like form factor of proton
and nuclei. Another source of uncertainty 
is the contribution coming from intermediate
states different then the proton or the nucleus in its ground state.
I has been proposed recently \cite{scholten} to
study the interference of virtual Compton scattering and the
Bethe-Heitler process as a way to extract nucleon properties, that
is, the form factor for the lepton pair production in virtual Compton
scattering. The result of such a measurement can be easily used to
improve our calculations.
\section{Acknowledgments}
This work was supported in part by the Swiss National Science
Foundation (SNF), the ``Freiwillige Akademische Gesellschaft''
(FAG) of the University of Basel, and the ``Deutsche 
Forschungsgemeinschaft'' (DFG). 
One of us (K.H.) would like to thank them for their financial support.

\begin{table}
\caption{The total cross section of electron-positron pair production by 
bremsstrahlung (brems. p.p) and in two-photon physics (two phot.phy. p.p.) 
at RHIC and LHC energies for colliding protons, calcium, lead or gold nuclei.}
\label{brems-tpp_e}
\begin{tabular}{l c c c c}
&$\gamma_{collider}$&$\sigma_{e^-e^+}^{brems.\ p.p.}(barn)$&
$\sigma_{e^-e^+}^{two\ phot.\ phy.\ p.p.}(barn)$&$\frac{\sigma_{e^-e^+}
(brems.\ p.p.)}{\sigma_{e^-e^+}(two\ phot.\ phy.\ p.p.)}$\\
\hline 
gold&100&$3.08*10^{-4}$&$2.74*10^4$&$1.1*10^{-8}$\\
proton (w/o VMD)&200&$3.52*10^{-11}$&$1.33*10^{-3}$&$2.7*10^{-8}$\\
lead&2900&$8.73*10^{-4}$&$2.05*10^5$&$4.3*10^{-9}$\\
calcium&3700&$4.33*10^{-6}$&$8.37*10^{2}$&$5.2*10^{-9}$\\
proton&7500&$5.88*10^{-11}$&$6.69*10^{-3}$&$8.8*10^{-9}$\\
proton (VMD)&7500&$5.88*10^{-11}$&$6.69*10^{-3}$&$8.8*10^{-9}$\\
\end{tabular}
\end{table}
\begin{table} 
\caption{(as TABLE I, but for muon-antimuon pair production)}
\label{brems-tpp_m}
\begin{tabular}{l c c c c}
 &$\gamma_{collider}$&$\sigma_{\mu_-\mu_+}^{brems.\ p.p.}(barn)$&$\sigma_
{\mu^-\mu^+}^{two\ phot.\ phy.\ p.p.}(barn)$&$\frac{\sigma_{\mu^-\mu^+}
(brems.\ p.p.)}{\sigma_{\mu^-\mu^+}(two\ phot.\ phy.\ p.p.)}$\\
\hline
gold&100&$4.77*10^{-5}$&$4.66*10^{-1}$&$1.0*10^{-4}$\\
proton&200&$1.42*10^{-12}$&$3.14*10^{-8}$&$4.5*10^{-5}$\\
proton (VMD)&200&$2.8*10^{-10}$&$3.14*10^{-8}$&$2.7*10^{-3}$\\
lead&2900&$2.02*10^{-4}$&$4.60*10^{0}$&$4.4*10^{-5}$\\
calcium&3700&$7.76*10^{-7}$&$1.78*10^{-2}$&$4.4*10^{-5}$\\
proton&7500&$2.80*10^{-12}$&$1.36*10^{-7}$&$2.0*10^{-5}$\\
proton (VMD)&7500&$5.7*10^{-10}$&$1.36*10^{-7}$&$1.2*10^{-3}$\\
\end{tabular}
\end{table}
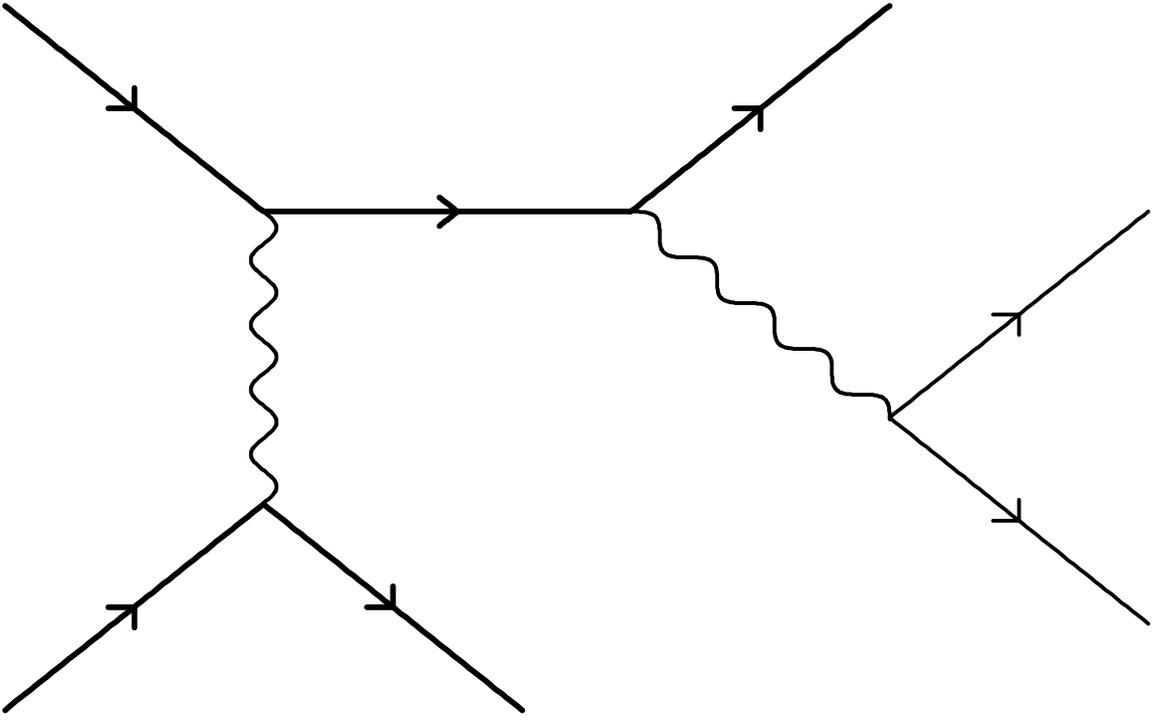
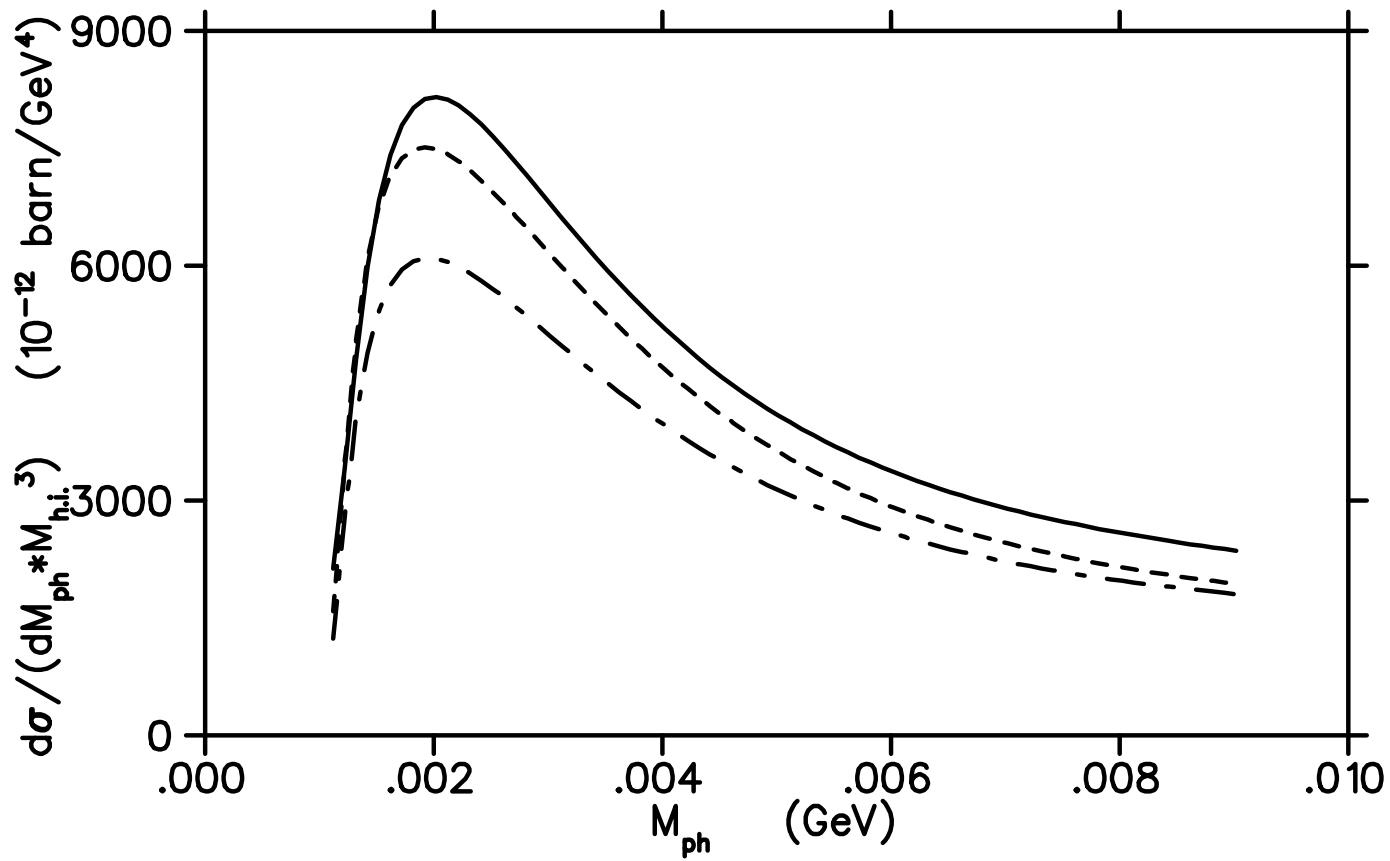
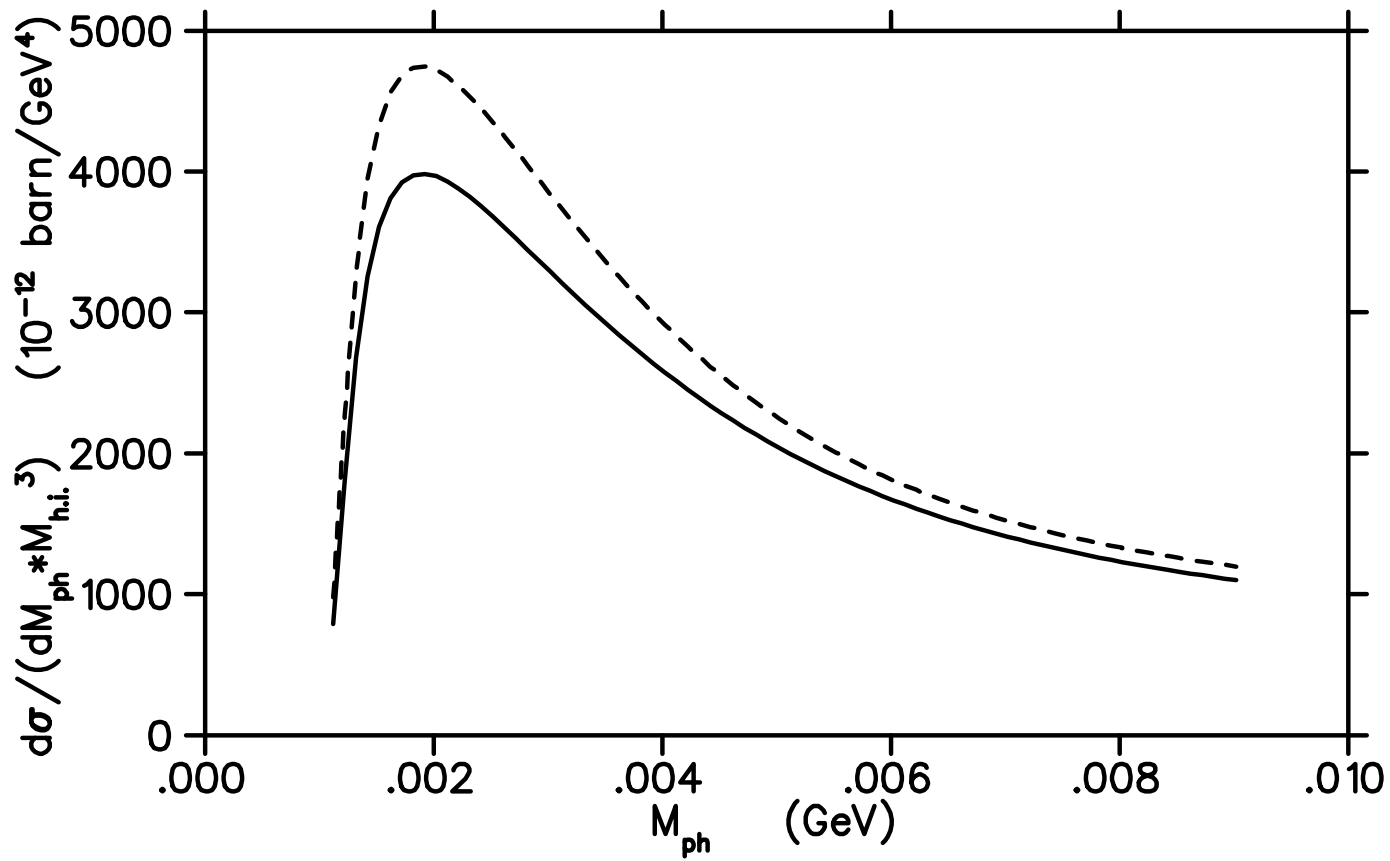
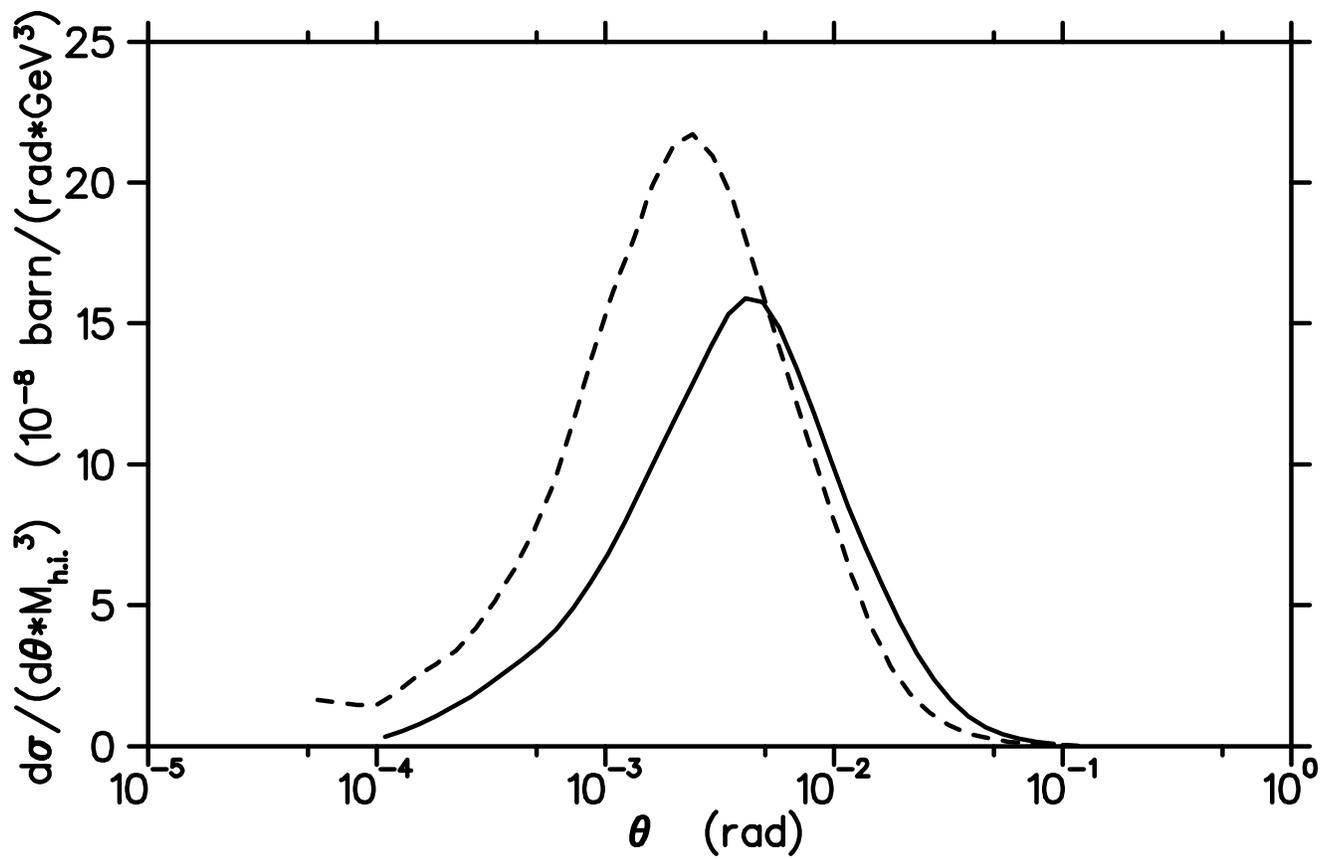
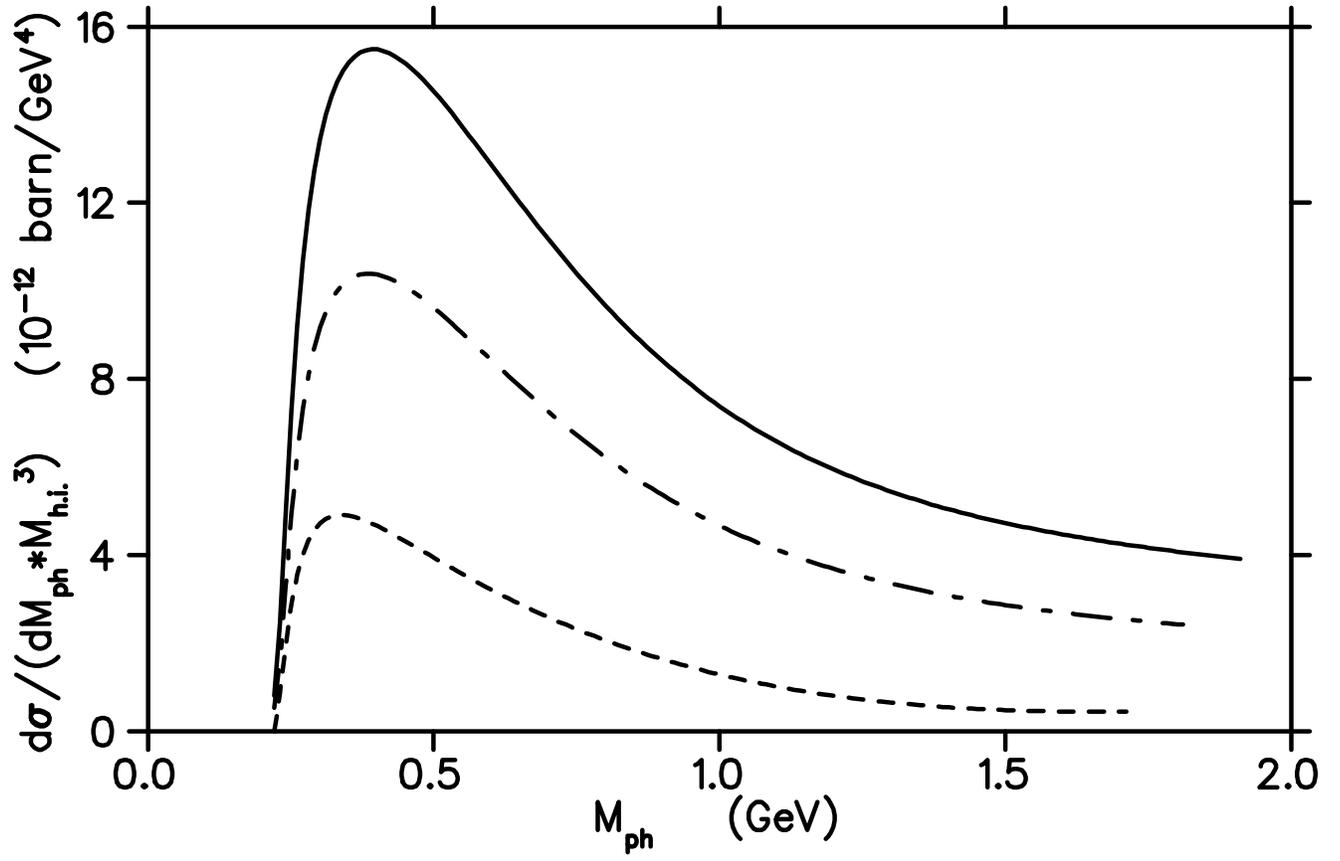
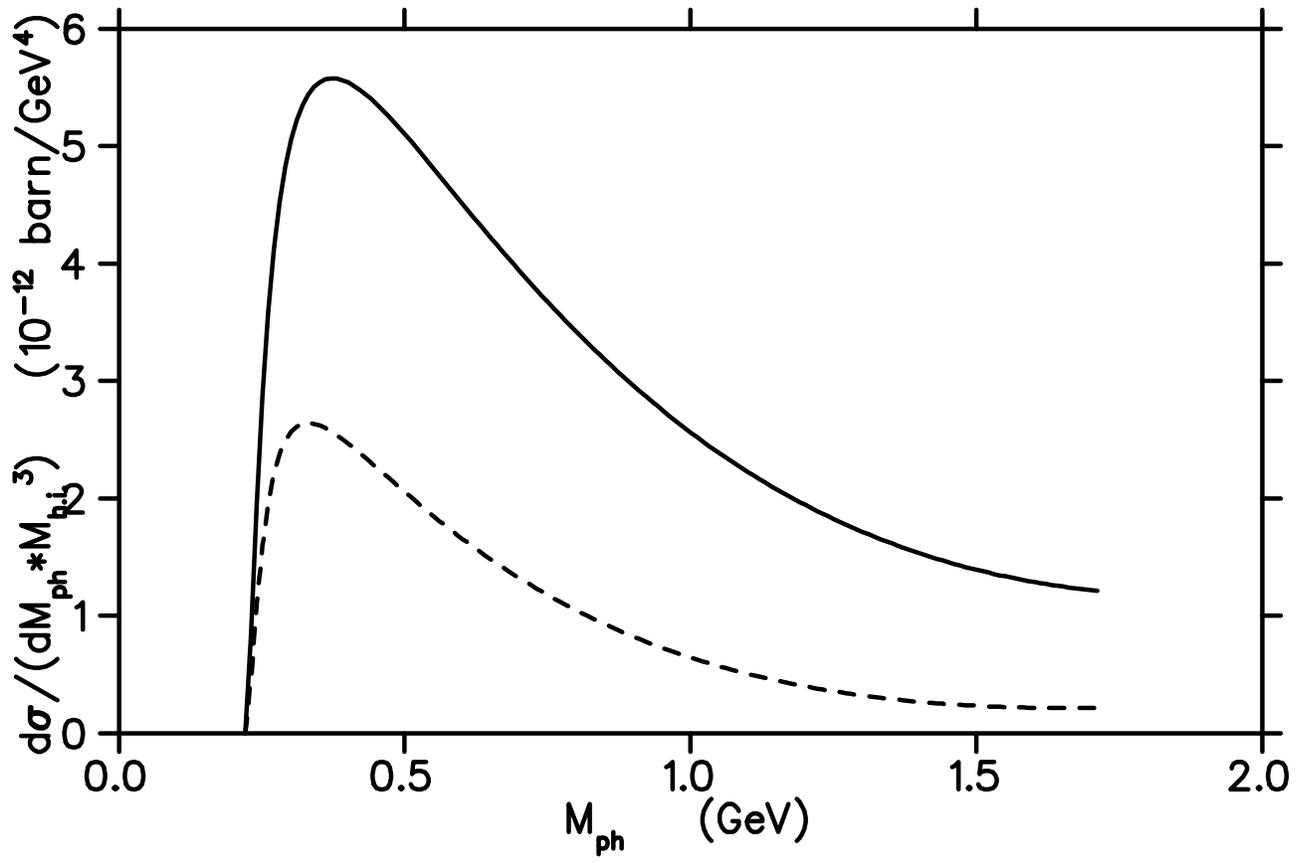
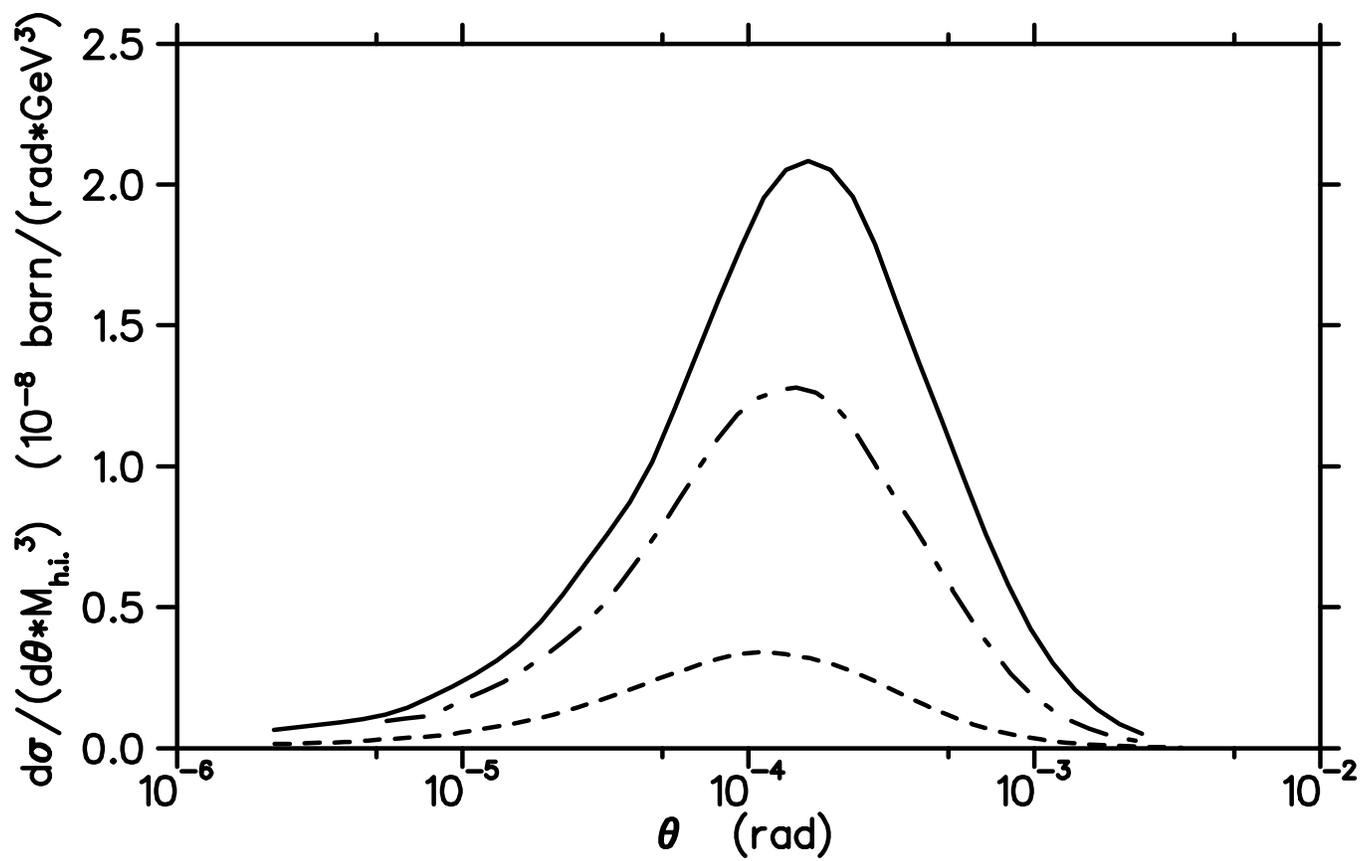
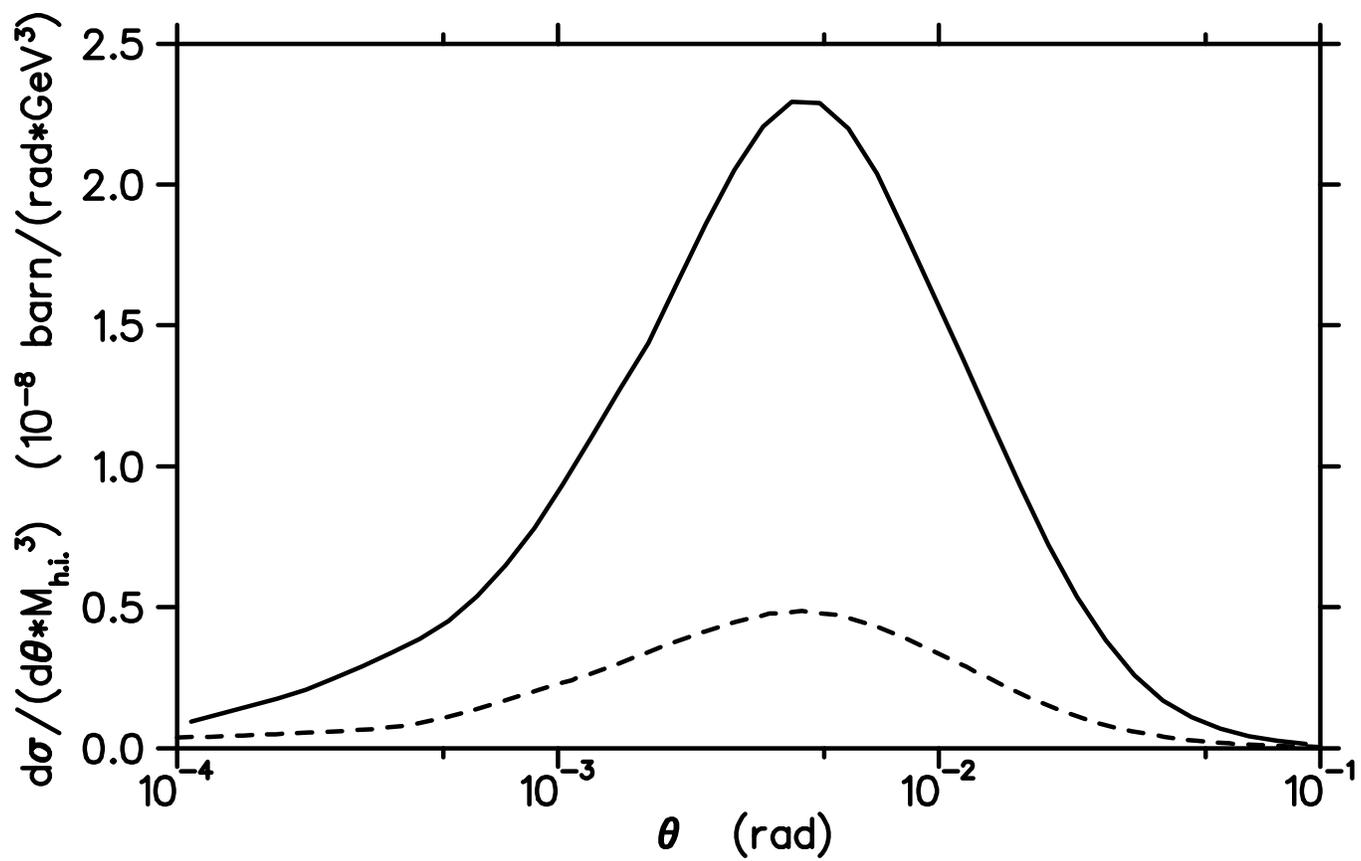
\end{document}